\documentclass[manuscript]{aastex}
\usepackage{amssymb}
\usepackage{amsmath}
\usepackage[colorlinks,linkcolor=blue,anchorcolor=blue,citecolor=blue,urlcolor=blue,dvipdfm]{hyperref}
\newcommand{\ie}{i.\,e.}

\shorttitle{Quantum simulations on transport properties of He}
\shortauthors{Kang et al.}

\begin{document}

\title{Quantum path integral molecular dynamics simulations on transport properties of dense liquid helium}

\author{Dongdong Kang, Jiayu Dai, Huayang Sun}
\affil{Department of Physics, College of Science, National
University of Defense Technology, Changsha 410073, Hunan, People's
Republic of China} \email{jydai@nudt.edu.cn}

\and

\author{Jianmin Yuan}
\affil{Department of Physics, College of Science, National
University of Defense Technology, Changsha 410073, Hunan, People's
Republic of China}
\affil{IFSA Collaborative Innovation Center,
Shanghai Jiao Tong University, Shanghai 200240, People's
Republic of China}
\email{jmyuan@nudt.edu.cn}

\begin{abstract}
Transport properties of dense liquid helium under the conditions of
planet's core and cool atmosphere of white dwarfs have been
investigated by using the improved centroid path-integral
simulations combined with density functional theory. The
self-diffusion is largely higher and the shear viscosity is notably
lower predicted with the quantum mechanical description of the
nuclear motion compared with the description by Newton equation. The
results show that nuclear quantum effects (NQEs), which depends on
the temperature and density of the matter via the thermal de Broglie
wavelength and the ionization of electrons, are essential for the
transport properties of dense liquid helium at certain astrophysical
conditions. The Stokes-Einstein relation between diffusion and
viscosity in strongly coupled regime is also examined to display the
influences of NQEs.
\end{abstract}

\keywords{dense matter - dense helium - diffusion - viscosity -
quantum effects}

\section{Introduction}
Helium, the second most abundant element in the universe, is one of
the main components of giant planets and a large number of recently
discovered exoplanets \citep{Santos05,McMahon12}. Meanwhile, helium
is also expected to be largely present in the cool white dwarf
atmospheres \citep{Winisdoerffer05}, which generally affect the
cooling rate of white dwarfs (WDs) when having exhausted their
nuclear fuel. Matter in these astrophysical bodies is generally
under extreme temperature and pressure conditions
\citep{Chabrier07}. For instance, the temperature range of Jupiter,
a typical solar giant planet, is estimated from 6500 K to 20000 K,
while the pressure can reach 70 Mbar \citep{Fortov11}. For some
exoplanets, it can be up to 19 Gbar \citep{Swift12}. The density
range of the outer layer of WDs is a few g/cm$^3$ and the
temperature retains thousands of Kelvin \citep{Iglesias02}. The
central element for determining the structure and evolution of these
astrophysical objects is the accurate description of the physical
properties of matter under the relevant conditions of temperature
and pressure \citep{Fortney10,Kang11,Kang13,French12,Militzer13},
which includes transport properties of the characteristic materials.
In fact, atomic diffusion, as a basic physical process, plays a
significant role in modeling the interior evolution of astrophysical
objects \citep{Michaud07}. Therefore, the physical properties of
helium under extreme conditions is of continuous and strong interest
owing to the aforementioned wealth applications to astrophysics
\citep{Fortov11}.

Despite the simplicity of atomic structure, bulk helium exhibits
complex behavior, especially under astrophysical conditions because
the states cover from atomic, partially ionized to fully ionized
helium \citep{Kietzmann07,Khairallah08,Stixrude08,McMahon12}. In
past few decades, great progresses have been achieved to explore the
physics of helium at extreme states experimentally, such as diamond
anvil cell and shock wave \citep{Celliers10,Soubiran12}. However,
only narrow bands of temperature and pressure parameters attainable
by laboratory measurements today. Thus the comprehensive
understanding of helium at extreme states requires accurate
theoretical calculations. Conventional approaches that are suitable
for treating ideal plasma gas or condensed matter under ambient
conditions usually fail to solve the issues in the warm dense regime
\citep{Graziani14}. Ab initio approaches based on density functional
theory (DFT) provide a feasible and satisfactory tool to investigate the physical
properties of matter under extreme conditions. They have been
extensively used to obtain the thermodynamical, transport, and
optical properties of matter in astrophysical area \citep{French12}.
Quantum Langevin molecular dynamics (QLMD), including non-adiabatic effects such as the
electron-ion collision induced friction in molecular dynamics
\citep{Dai10a}, extends the quantum molecular dynamics to the regime
up to solar interior \citep{Dai10b,Dai12,Dai13}.

A common feature of these theoretical methods is that the ions are
treated classically, that is, the quantum nature of ions motion is
neglected. It should be noted that due to the low mass, helium
motion is expected to be fairly sensitive to quantum effects, the
so-called nuclear quantum effects (NQEs), as hydrogen exhibits
\citep{Kang14}. The well-known superfluidity of helium is dominated
by quantum effects of helium atoms at extremely low temperatures
\citep{Ceperley95}. At relatively high temperatures in planet's core
or outer layer of WDs, NQEs are usually considered to be negligible
before. Note that with the increasing density of helium, the atomic
distance decreases. When the nearest atomic distance is comparable
to the thermal de Broglie wavelength, the nuclear quantum
delocalization would play remarkable roles in structure and
transport properties, e.g., NQEs induce complex transport behavior
in warm dense hydrogen as reported in our previous work
\citep{Kang14}. Path-integral molecular dynamics is an effective
tool to investigate the NQEs on static properties of extreme states
\citep{Marx96} and recently improved centroid path-integral
molecular dynamics (PIMD) can provide the well-defined
quasiclassical approximation of real-time evolution of quantum
systems \citep{Cao93,Kang14}. Therefore, centroid PIMD,
combined with DFT, treats both electrons and ions
quantum-mechanically. Using this method, more accurate description
of transport properties of matter at extreme states is expected to
be obtained.

In this study, we investigate NQEs on the structural and transport
properties of dense liquid helium at 5 g/cm$^3$ and 10 g/cm$^3$
using ab initio centroid PIMD with an improved scheme, and compare
the results with that from traditional first principles molecular
dynamics (MD) calculations based on DFT with a classical treatment
of nuclei to assess the importance of NQEs.

\section{Theoretical Methods}
Path-integral formulation of quantum statistical mechanics provides
a theoretical and computational framework to study the quantum
many-body system at finite temperatures \citep{Feynman72}. In
path-integral picture, the canonical partition function can be
expressed as a configurational integral of Boltzmann weighted
continuous paths. If exchange effects of particles are negligible,
these closed paths can be discretized through Trotter approximation
into $P$ beads circularly connected via harmonic springs, so that a
quantum system including $N$ particles is isomorphic to a classical
system consisting of $N$ ring polymers, where each quantum particle
is mapped onto a closed flexible polymer of $P$ beads
\citep{Parrinello84,Marx96}. Static properties, e.g., radial
distribution function (RDF), of the quantum system are routinely
obtained through molecular dynamics sampling the configuration space
of its isomorphic classical system. Further, centroid molecular
dynamics within the framework of path-integral improved recently is
a good quasiclassical approximation to quantum dynamics
\citep{Kang14}. In the adiabatic centroid PIMD scheme
\citep{Marx99}, the primitive path variables
$\{\boldsymbol{R}_I\}^{(s)}$ ($s$ is the imaginary time.) is
transformed to a set of normal mode variables
$\{\boldsymbol{y}_I\}^{(s)}$, which diagonalize the harmonic bead
coupling. The equations of motion of normal mode variables without
thermostats are given by
\begin{align}
M'^{(1)}_I\ddot{\boldsymbol{y}}^{(1)}_I &=-\frac{\partial}{\partial \boldsymbol{y}^{(1)}_I}\frac{1}{P}\sum_{s=1}^{P}E^{(s)}, \\
M'^{(s)}_I\ddot{\boldsymbol{y}}^{(s)}_I &=-\frac{\partial}{\partial \boldsymbol{y}^{(s)}_I}\frac{1}{P}\sum_{s'=1}^{P}E^{(s')}-M^{(s)}_I\omega^2_P\boldsymbol{y}^{(s)}_I, \quad s=2,\cdots, P,
\end{align}
where $\omega_P=\sqrt{P}k_BT$, $P$ is the number of beads in ring
polymer, $k_B$ is Boltzmann constant, $T$ is the target temperature
of the simulated system, $E^{(s)}$ is the electronic energy
functional in \textit{ab initio} calculations.

The centroid and non-centroid modes are mass-scale separated with
adiabatic parameter $\gamma$, \ie, the normal mode masses
$M'^{(s)}_I$ are
\begin{equation}
M'^{(1)}_I=M_I,\quad M'^{(s)}_I=\gamma M^{(s)}_I, s=2,\cdots,P.
\end{equation}
where $M_I$ is the physical nuclear mass. In this way, the nuclear
quantum dynamics is obtained in the quasiclassical approximation
through driving the centroid move in real-time in the centroid
effective potential generated by all non-centroid modes. Here we
propose a new scheme for sampling the force on the centroids during
the molecular dynamics simulations to make the size of ring polymer
is more close to the de Broglie wavelength as the instantaneous
kinetic energy of ions is decreased. That is, for each ion
$\omega_P$ depend linearly on its instantaneous kinetic energy
$E(t)$ during simulations,
\begin{equation}
\omega_P(t)=\frac{2N}{g}\sqrt{P}(aE(t)+bE_0),
\end{equation}
where $t$ is the simulation time step. The details of determining
the parameters $a$ and $b$ are given otherwhere \citep{Kang14}. Here
we show the results, i.e., $a=1-(\frac{2R_{g0}}{\lambda})^2$ and
$b=(\frac{2R_{g0}}{\lambda})^2$, where $R_{g0}$ is the radius of
gyration of ring polymer when $E(t)=E_0$, $\lambda$ is the de
Broglie wavelength of the corresponding ion. As shown in
\citet{Kang14}, this improved centroid PIMD approach can describe
the quantum collision properly, and nuclear quantum dynamics in
dense hydrogen has been investigated successfully.

\section{Results and Discussion}
\subsection{Computational details}
For dense helium, we study the structural and transport properties
at densities of 5 g/cm$^3$ and 10 g/cm$^3$ above the melting point
temperature. The knowledge of the structure and transport properties
of dense liquid helium under these density and temperature
conditions is essential to correctly describe the interior structure
and evolution and cooling of astrophysical bodies, such as
exoplanets and outer layer of white dwarfs. Exchange effects between
ions are neglected in this study \citep{Dyugaev90} because the
atomic distance is still larger than the thermal de Broglie
wavelength of ions under the conditions considered here.

Our MD and PIMD simulations were performed using the modified
Quantum-ESPRESSO package \citep{QE}, where electrons are
quantum-mechanically treated by finite-temperature density
functional theory \citep{Mermin65}. The interaction between valence
electron and the ionic core is presented by norm-conserving
pseudopotential. The generalized-gradient approximation for
exchange-correlation potential \citep{PBE} is adopted. The
plane-wave kinetic energy cutoff was set from 150 to 180 Ry
according to different densities. The Brillouin zone is sampled
using the Gamma point in dynamic simulations, while denser k-point
grids of 4$\times$4$\times$4 Monkhorst-Pack scheme points are used
to calculate electronic properties. The periodic supercell including
128 or 250 atoms is adopted for 5 g/cm$^3$ and 10 g/cm$^3$, which
can ensure convergence of both ionic and electronic properties with
good accuracy.

Within the framework of PIMD, the structural properties are
calculated using the primitive scheme, while the real-time quantum
dynamics of nuclei is obtained through the improved centroid PIMD.
The self-diffusion coefficient is calculated from the slope of the
centroid mean square displacement, while the shear viscosity is
obtained from the autocorrelation function of the off-diagonal
components of the stress tensor \citep{Alfe98}. Langevin thermostat
is employed to overcome the nonergodic problem, which not only
produces a canonical ensemble and compensates the calculated errors,
but also gives an good method to include the non-adiabatic effects of
electrons in warm and hot dense regime \citep{Dai10a,Dai10b}. Note
that Langevin thermostat is applied only to each noncentroid degree
of freedom because large thermostats would disturb the dynamical
properties of the centroids. The Trotter number was set to 16 after
a convergence test. Time step of 7-9 a.u. is used in the MD
calculations with 30000 steps after thermalization, while the
smaller time steps of 0.3 a.u. with 5$\times$10$^5$ steps and a
centroid adiabaticity parameter of 20 are adopted in the centroid
PIMD simulations in order to decouple the centroid mode from other
normal modes.

\subsection{Transport properties of dense liquid helium}
Firstly, we calculated the RDF of helium nuclei to reveal the
structural difference introduced by NQEs. Within the framework of
imaginary-time PIMD, the quantum expectation value of a static
physical quantity is given in terms of averaging over the canonical
ensemble of the isomorphic classical system. Therefore, the radial
distribution functions of helium atoms are obtained by averaging
over both the PIMD time steps and imaginary time slices. Figure 1
shows that NQEs play an important role in RDF at low temperatures
with the density range of 5 g/cm$^3$ and 10 g/cm$^3$. In particular,
the first peak of RDF from PIMD simulations are largely reduced and
broadened, which indicates the significant nuclear quantum
delocalization. When the temperature is increased, the nuclear
quantum delocalization becomes weaker and cannot be observed
ultimately at 10000 K for 5 g/cm$^3$ helium and 12000 K for 10
g/cm$^3$ helium. Meanwhile, RDF shows more pronounced nuclear
quantum effect with increasing density, even though the
temperature (7000 K) at 10 g/cm$^3$ is higher than that (4000 K) at
5 g/cm$^3$.

Of our particular interest is NQEs on transport properties of dense
liquid helium because atomic diffusion is the basic physical process
that determine the interior structure and evolution of astrophysical
objects \citep{Michaud07}. The results of the self-diffusion
coefficient and shear viscosity at 5 g/cm$^3$ with the temperature
range of 4000-10000 K are shown in Figure 2. When taking into account
the quantum features of the collision processes between ions
properly, the self-diffusion coefficients obtained by centroid PIMD
simulation are substantially larger than the classical simulation
value over the whole temperature range of 4000-10000 K. The
difference is 37$\%$ at 4000 K, 22$\%$ at 6000 K, 17$\%$ at 8000 K
and 10$\%$ at 10000 K. It is interesting that even though the RDF
from MD and PIMD calculations are almost identical when the
temperature is increased to 10000 K, the ionic diffusion still
exhibits distinct difference with the inclusion of NQEs. In fact,
the essence of atomic diffusion is atomic scattering from the
viewpoint of collision physics. The large-angle scattering between
ions is dominant under the high density conditions considered here.
The pronounced quantum nuclear scattering features introduced by
centroid PIMD simulations leads to a smaller large-angle scattering
cross section between ions than the classical treatment, thereby
increasing the mean free path of ions. Thus the classical-particle
treatment of protons substantially underestimates the ionic
diffusion. Similarly, the self-diffusion coefficients at the
temperature range of 7000-12000 K from quantum simulations for
helium at 10 g/cm$^3$, as shown in Figure 3, are 23 $\%$, 21 $\%$,
12 $\%$ and 6 $\%$ higher than that from classical simulations.

The shear viscosities calculated from the Green-Kubo relation
\citep{Alfe98} at the density of 5 g/cm$^3$ and 10 g/cm$^3$ are also
shown in Figure 2 and 3, respectively. We can see that the
viscosities from centroid PIMD are smaller than those from MD and
the difference between them becomes smaller with increasing
temperatures. In addition to obtaining the viscosity from molecular
dynamics simulations, it can be alternatively deduced from the
Stokes-Einstein (SE) relation with slip or stick boundary conditions
\citep{Hansen11}. The SE relation, which is exact for the Brownian
particles, is not valid in strongly coupled regime. In order to
examine the SE relation, we calculated the parameter $D\eta a/k_BT$
(a is the effective atomic diameter and defined by the position of
the first peak of RDF) and the results are shown in Figure 2 and 3
for the density of 5 g/cm$^3$ and 10 g/cm$^3$, respectively. The
parameter $D\eta a/k_BT$ both from MD and PIMD calculations deviates
from the SE relation value $1/2\pi$ at low temperatures;
furthermore, NQEs introduce pronounced difference between the
classical and quantum simulations. In fact, the state of helium in
planet's core or cool atmosphere of WDs is in moderately or strongly
coupled regime and the motion of ions is quite different from
Brownian motion.

In order to understand the atomic diffusion behavior with NQEs microscopically, we explore the potential surface of a randomly selected atom along its simulation trajectory. Given the force in simulations is calculated by $\boldsymbol{F}=-\nabla V$, the potential energy along the trajectory of a atom can be obtained by $V=V_0+\int \boldsymbol{F} \cdot d\boldsymbol{r}$, where $\boldsymbol{F}$ is the force acting on atoms, $V$ is the potential energy, $V_0$ is the reference point of potential energy. The potential energy of a randomly selected helium atom along its simulation trajectory from MD and PIMD simulations as well as the distribution of the potential energy is presented in Figure 4. We can see that in PIMD simulations the potential energy of the imaginary-time beads fluctuates steeply, which presents the effects of quantum fluctuations on dynamics of the centroid. The potential surface of the centroid exhibits more smoothly than that of classical particle, which is evidenced by the fact that the distribution of the potential energy of centroid covers smaller energy range than MD simulation and the high frequency component of the potential surface of the centroid is higher than MD results by about 100 THz for both 5 g/cm$^3$ and 10 g/cm$^3$, as shown in Figure 4(c) and (f). It indicates that when considering NQEs the atoms are more prone to turn over the energy barrier than classical treatment, thereby enhancing the atomic diffusion. In addition, it is shown in Figure 4 that when the density is increased from 5 g/cm$^3$ to 10 g/cm$^3$, the potential surface exhibits more barriers and wells in the same range of time (100 fs) both in MD and PIMD simulations. It unambiguously indicates that helium atoms collide more frequently at higher density at the same temperature so that the atomic diffusion is lowered at high densities.

Nuclear motion quantum nature not only affect the atomic diffusion
behavior, but also induce the electronic redistribution \citep{Cannuccia11}, thus
introducing more localized electrons \citep{Kang13,Kang14}. It can
be evidenced by the distribution of charge density shown in Figure
5. Note that with the inclusion of NQEs, the charge density has the
larger maxima and smaller minima compared with the classical
simulations, which indicates that electrons surrounding ions distribute
more localized with quantum nuclei than classical particle
treatment. For metallic state of matter, e.g., dense hydrogen, this
electronic localization can significantly lower the dc electrical
and thermal conductivities \citep{Kang14}. For dense liquid helium, NQEs induce a strong impact on the electronic density of states (DOS). We can see from Figure 6 that the energy bands, especially conduction bands are obviously broadened when NQEs are considered via PIMD. Since helium exhibits a large electronic energy gap under the conditions considered here, it is still in insulator state. Therefore, NQEs on electronic transport properties are not discussed here. Except for NQEs on DOS, density effect has also influenced the profile of DOS remarkably. As shown in Figure 6, when the density of helium is increased from 5 g/cm$^3$ to 10 g/cm$^3$, the energy gap is decreased, thus the \textit{K}-edge photo-absorption spectra will have a red shift. More importantly, we note that the DOS of the 1s state is extended by about 30 eV when helium is compressed from 5 g/cm$^3$ to 10 g/cm$^3$, which significantly affects the interaction potential between helium nuclei and make the potential surface fluctuates more frequently at high density (shown in Figure 4). Therefore, both NQEs and density effect contributes to the complex transport properties of atoms.

\section{Conclusion}
In conclusion, NQEs on transport properties of dense liquid helium
under the conditions of the planet's core and cool atmosphere of WDs
have been investigated using the improved centroid PIMD simulations
combining with DFT. The results show that with the inclusion of NQEs,
the self-diffusion is largely higher while the shear viscosity is
notably lower than the results of without the inclusion of NQEs due to the lower collision cross sections even when the NQEs have little effects on the static structures. Meanwhile,
the relation between diffusion and viscosity is deviate from the SE
relation valid for Brownian particles, thus the SE relation is not
valid in strongly coupled regime. The potential surface of helium atom along the simulation trajectory is quite different between MD and PIMD simulations. We have shown that the quantum nuclear character induces complex behaviors for ionic transport
properties of dense liquid helium. Therefore, in order to construct
more reasonable structure and evolution model for the planets and
WDs, NQEs must be reconsidered when calculating the transport
properties at certain temperature and density conditions.

\acknowledgments

This work is supported by the National Basic Research Program of
China (973 Program) under Grant No. 2013CB922203, the National
Natural Science Foundation of China under Grant Nos. 11422432,
11274383 and 61221491. DJY thanks the support by the Advanced Research Foundation of National University of Defense Technology. Calculations were carried out at the Research
Center of Supercomputing Application, NUDT.

\clearpage

\begin{figure}
\epsscale{.70} \plotone{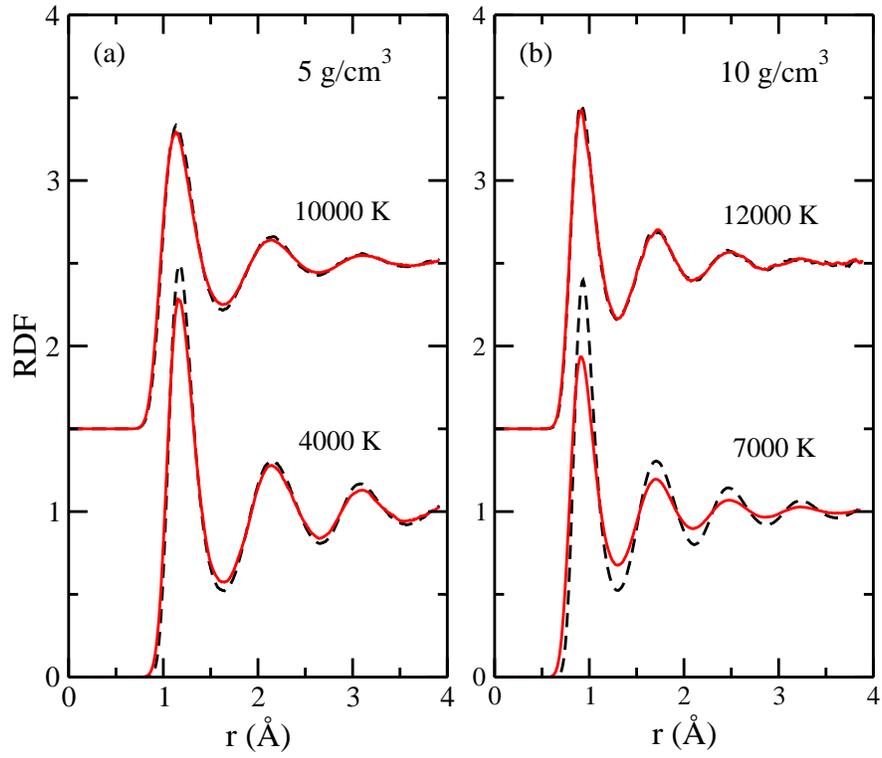} \caption{Radial distribution
functions of helium nuclei from PIMD (solid lines) and MD (dashed
lines) simulations with different temperatures at 5 g/cm$^3$ (a) and
10 g/cm$^3$ (b).\label{fig1}}
\end{figure}

\clearpage

\begin{figure}
\epsscale{.60} \plotone{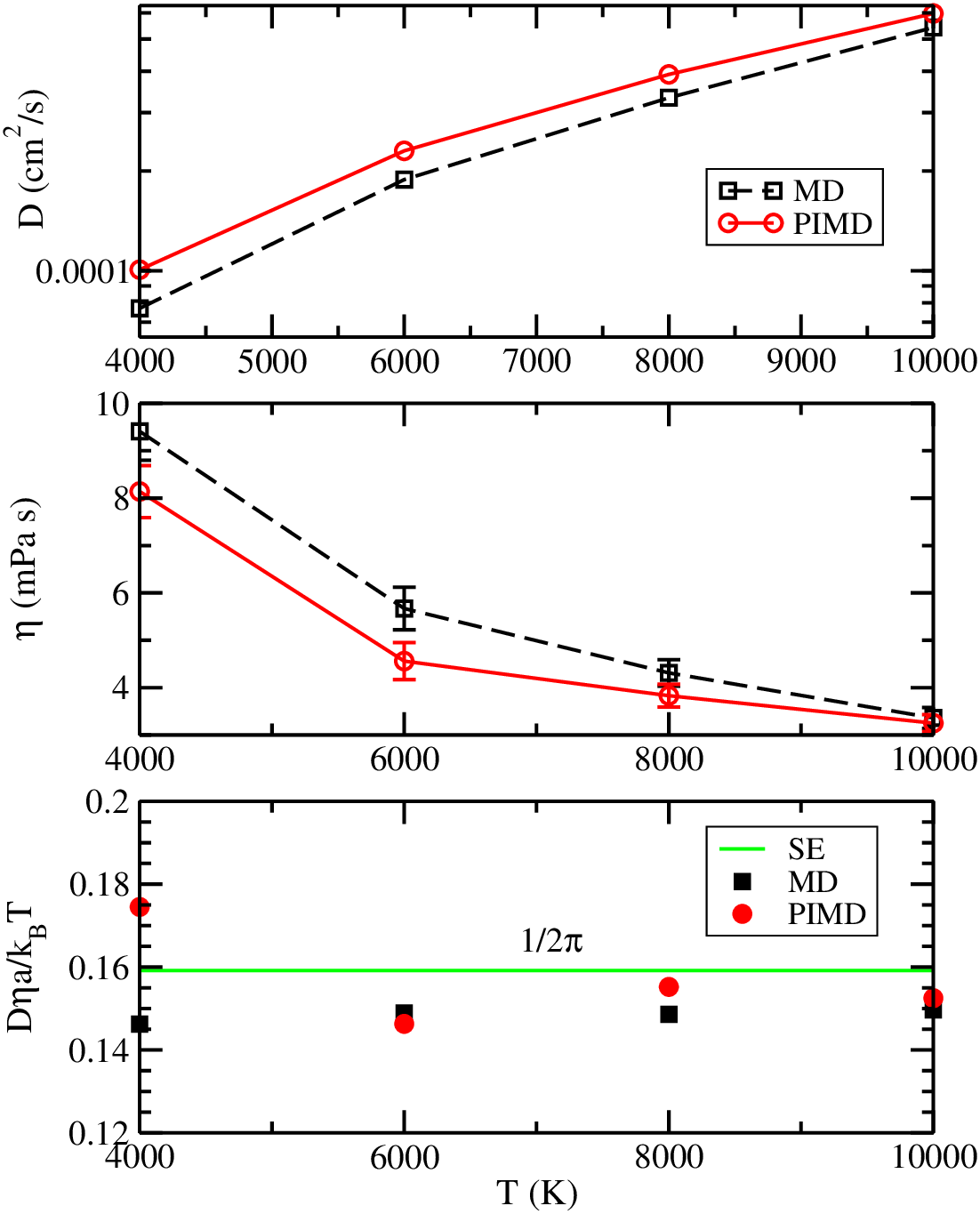} \caption{Temperature dependence of
the self-diffusion coefficient and shear viscosity at 5 g/cm$^3$
with temperatures of 4000 K, 6000K, 8000 K and 10000 K. obtained
from centroid PIMD and MD simulations. The shear viscosities
obtained from the Stokes-Einstein (SE) relation are also presented.
The effective atomic diameter in SE relation is defined by the
position of the first peak of radial distribution
function.\label{fig2}}
\end{figure}

\clearpage

\begin{figure}
\epsscale{.60} \plotone{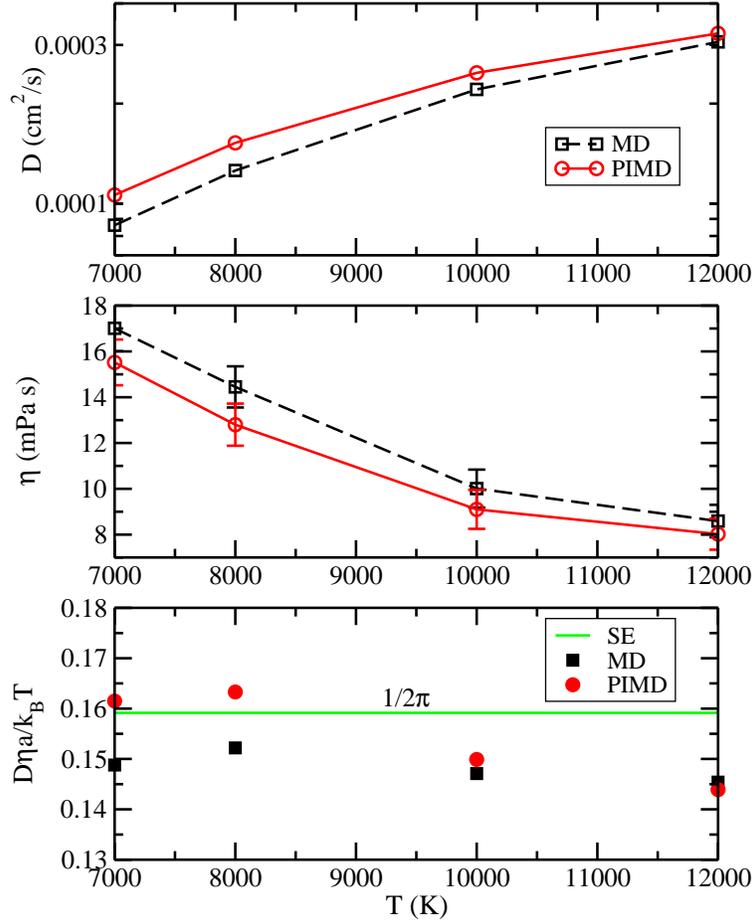} \caption{Temperature dependence of
the self-diffusion coefficient and shear viscosity at 10 g/cm$^3$
with temperatures of 7000 K, 8000K, 10000 K and 12000 K. obtained
from centroid PIMD and MD simulations. The shear viscosities
obtained from the Stokes-Einstein (SE) relation are also presented.
The effective atomic diameter in SE relation is defined by the
position of the first peak of radial distribution
function.\label{fig3}}
\end{figure}

\clearpage

\begin{figure}
\epsscale{.80} \plotone{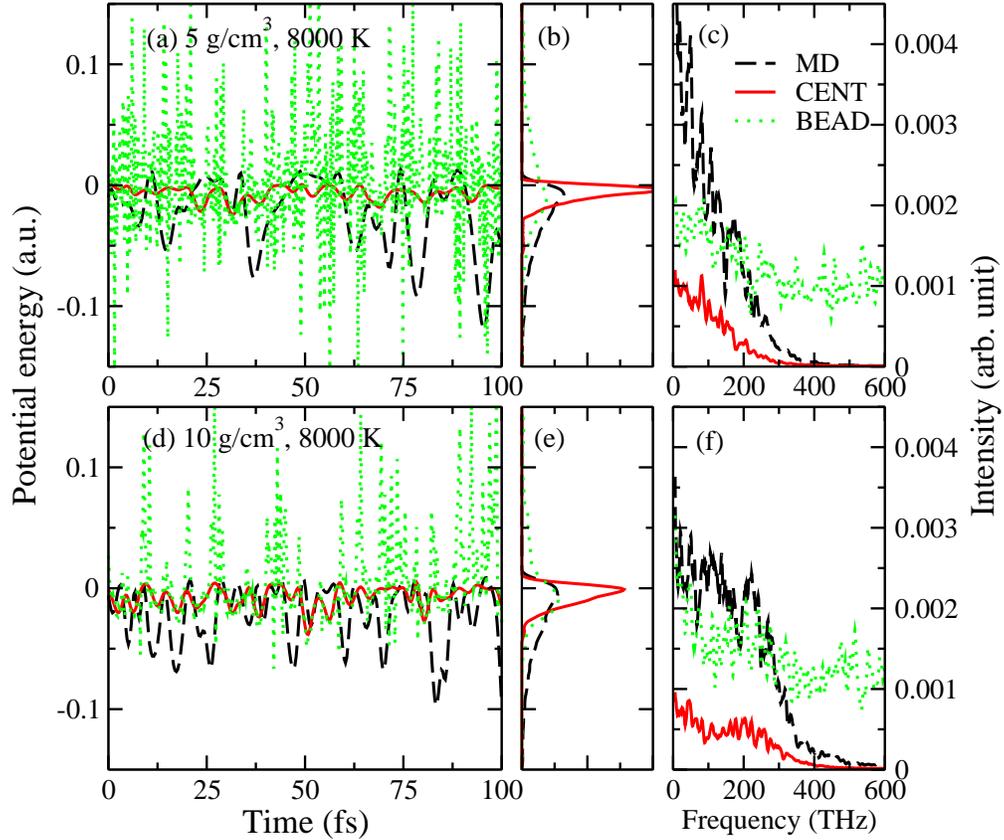} \caption{The potential energy of a randomly selected helium atom along its simulation trajectory. The results of classical simulation (dashed lines), the centroid of PIMD (solid lines) and the randomly-selected bead in ring polymer (dotted lines) are displayed in two conditions of (5 g/cm$^3$, 8000 K) and (10 g/cm$^3$, 8000 K). Middle panels (b) and (e) are the distribution of the potential energy. Note that the peaks of distributions are shifted to the same value for comparisons. Right panels (c) and (f) are the fourier transform of the potential surface.\label{fig4}}
\end{figure}

\clearpage

\begin{figure}
\epsscale{.80} \plotone{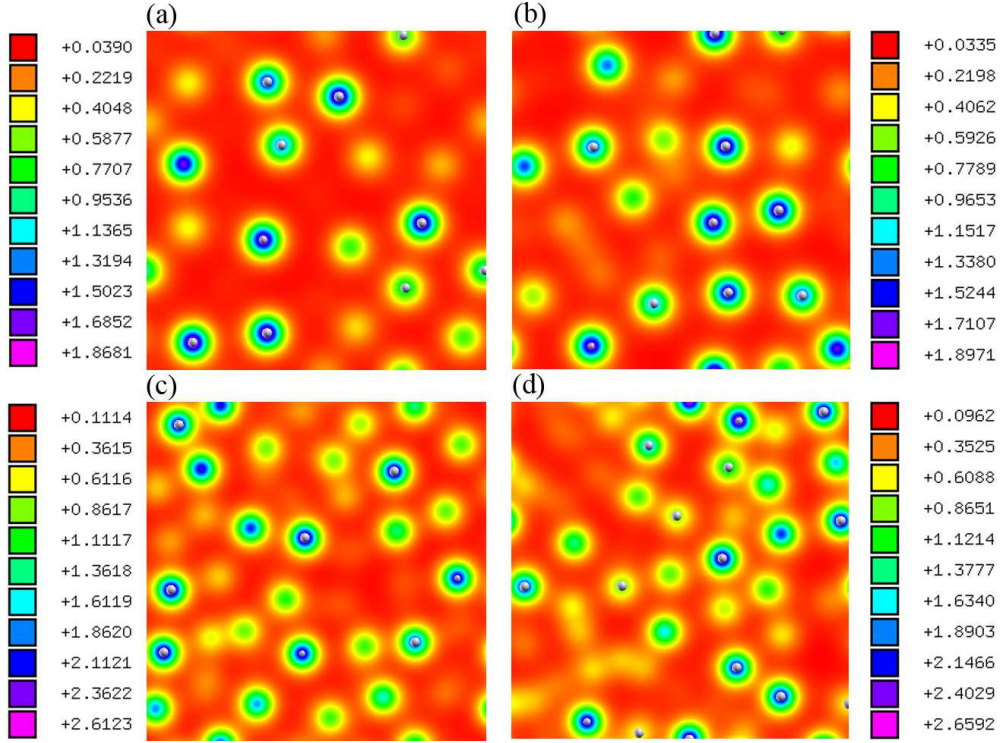} \caption{The cut-plane of 3D
distribution of charge density of randomly selected configurations
from MD (a, c) and PIMD (b, d) simulations. For upper panel, the density is 5
g/cm$^3$ and the temperature is 4000 K. For lower panel, the density is 10 g/cm$^3$ and the temperature is 7000 K. In PIMD
calculations, the configuration is randomly selected from
imaginary-time slices in simulation time steps.\label{fig5}}
\end{figure}

\clearpage

\begin{figure}
\epsscale{.70} \plotone{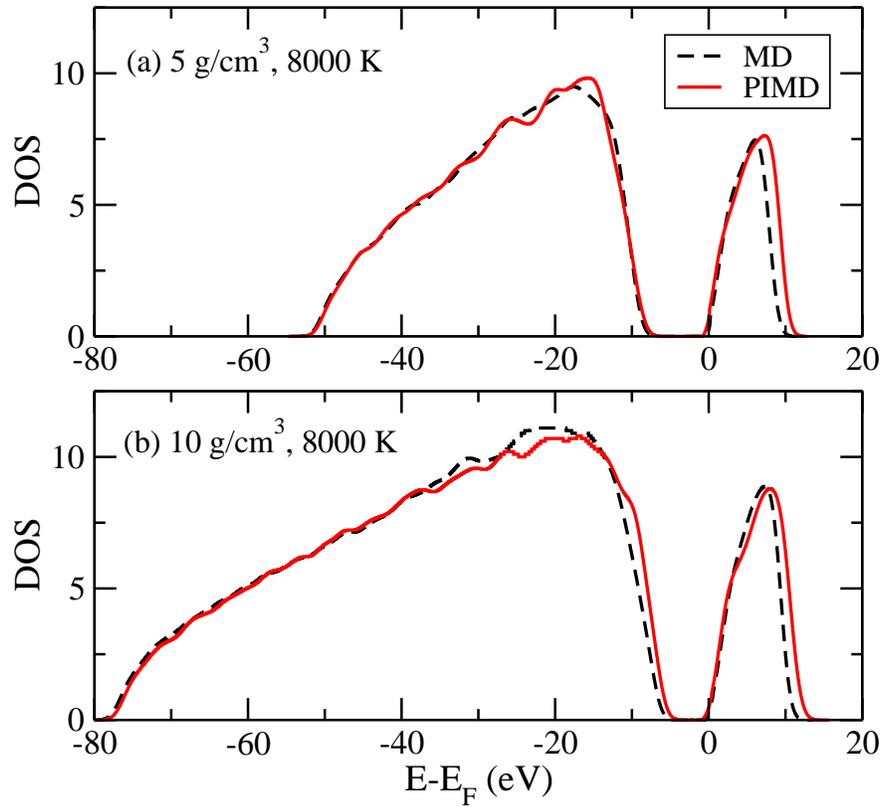} \caption{The electronic density of states (DOS) calculated from quantum (solid lines) and classical simulations (dashed lines) at 5 g/cm$^3$(a) and 10 g/cm$^3$(b). The results from PIMD are averaged over not only the simulation time step but also the imaginary-time slices in each simulation time step.\label{fig6}}
\end{figure}

\end{document}